\documentclass[floatfix,aps,10pt,prl,twocolumn,nopreprintnumbers,groupedaddress,superscriptaddress,tightenlines,longbibliography,letterpaper,altaffilletter,showpacs,nofootinbib]{revtex4-2}
\usepackage{amsmath,amsfonts,amssymb,amsthm,mathrsfs}
\usepackage{subfigure}
\usepackage{enumerate}
\usepackage{graphicx,physics,bm}

\usepackage[dvipsnames]{xcolor}
\RequirePackage{color}
\RequirePackage[colorlinks=true
,urlcolor=Maroon
,anchorcolor=Maroon
,citecolor=Maroon
,filecolor=Maroon
,linkcolor=blue
,menucolor=Maroon
,linktocpage=true
,pdfproducer=medialab
,pdfa=true
,pdfusetitle
]{hyperref}

\usepackage{natbib}
\usepackage{verbatim}
\usepackage[toc,page]{appendix}
\usepackage{epsf,bbm,amsbsy}
\usepackage{soul}
\usepackage{tikz}
\usetikzlibrary{patterns, positioning, arrows}
\usepackage{stackengine,scalerel}

\usepackage[T1]{fontenc}
\usepackage{times}

\allowdisplaybreaks
\definecolor{darkgreen}{RGB}{0,102,102}
\definecolor{purple}{RGB}{102,0,102}
\definecolor{darkblue}{RGB}{0,0,102}
\hypersetup{
    colorlinks=true,       
    linkcolor=purple,         
    citecolor=darkblue,        
    filecolor=magenta,     
    urlcolor=darkgreen         
}
\begin{document}

\title{Lorentzian Einstein-Ricci Flows}
\author{Aditya Dhumuntarao}
\affiliation{Perimeter Institute for Theoretical Physics, Waterloo, Ontario, N2J 2Y5, Canada}
\affiliation{School of Physics and Astronomy, University of Minnesota, Minneapolis, Minnesota, 55455, USA}
\email{dhumu002@umn.edu}

\begin{abstract}
	We study the Ricci flow for the Lorentzian Einstein-Hilbert action. We show that Einstein gravity emerges as a fixed point of the Einstein-Ricci flow equations and derive a renormalization group flow in Euclidean signature. By considering linearizations near the fixed point, the dynamics of the metric reveal that curvature deformations flow according to a forward heat equation with the stress-energy tensor acting as a source.
\end{abstract}
\maketitle
	The purpose of this note is to study the Ricci flow \cite{Hamilton:1982,2002math.....11159P,2003math......3109P,topping_2006} for the Lorentzian Einstein-Hilbert action
	\begin{equation}\label{eqn:EHaction}
		\mathcal{A}_L = \frac{1}{2\kappa}\int_{\mathcal{M}}{R}\sqrt{-g}\dd^d x
	\end{equation}
	where ${R}$ is the scalar curvature for a $d-$dimensional manifold $(\mathcal{M},g)$ without boundary $(d\ge3)$ and $\kappa=8\pi G_d$. Such geometric flows are interesting to consider as they connect on-shell saddle points of the action via deformations off-shell. The simplest example connects the four-dimensional Schwarzschild black hole to Minkowski spacetime in Euclidean signature \cite{Headrick:2006ti}. The transition between these geometries was shown to pass through a singularity in finite time. The Ricci flow has been studied several times -- originally in the context of non-linear sigma models \cite{PhysRevLett.45.1057}, and later as a geometrization of RG flows in AdS/CFT \cite{Woolgar:2007vz,Jackson:2013eqa,Figueras:2011va,Shyam:2018sro,Verlinde:1999xm}.

\section{Introduction}
	\label{sec:Introduction}
	For a $d$-dimensional manifold equipped with a Riemannian metric $(\mathcal{M},g)$, the Ricci flow is a first order PDE in the space of Riemannian metrics $\mathfrak{G}$ canonically given by
	\begin{equation}\label{eqn:RicciflowEqn}
		\dot{g}_{ab}(t) = - 2 R_{ab}(t),\hspace{.2in} g_{ab}(0)={g}^i_{ab}.
	\end{equation}
	The parameter $t$ appearing equation is a auxiliary time which characterizes the deformations of the metric along the flow. The Ricci flow is well defined in Euclidean signature where the PDE is parabolic (up to diffeomorphism invariance). The flow deforms the initial metric ${g}^i$ in the direction of its curvature and eventually terminates at $t^\star$ when the metric $g_{ab}(t^\star)$ is Ricci flat, $R_{ab}(t^\star)=0$. As a simple example, we can study the homothetic evolution of Einstein spaces ${R}_{ab}(t) = \frac{2\Lambda}{d-2} {g}^i_{ab}$ discussed earlier by Topping \cite{topping_2006}. In Lorentzian signature, this includes maximally symmetric spacetimes, Schwarzschild black holes $(\Lambda=0)$, and (A)dS black hole solutions $(\Lambda\ne0)$. A straightforward computation shows
	\begin{equation}
		g_{ab}(t) = \qty(1-\frac{4\Lambda}{d-2}t){g}^i_{ab}.
	\end{equation}
	Hence, spaces which are Ricci flat are fixed points whereas positively curved Einstein spaces $(\Lambda>0)$ become singular in finite time $t_c = \frac{(d-2)}{4\Lambda}$ and negatively curved Einstein spaces $(\Lambda<0)$ diverge homogeneously.

	In general, the Ricci flow describes a modified heat equation for Riemannian metrics. This is most easily seen for the flat space fixed point. First, consider a linearization of the flow around Ricci-flat space, i.e., $g_{ab}(t) = \bar{g}_{ab} + \epsilon h_{ab}(t)$. Then, up to linear order in $\epsilon$, the flow reduces to
	\begin{equation}
		\dot{h}_{ab} = -\bar{\Delta}_\text{L} h_{ab} + 2\bar{\nabla}_{(a}\xi_{b)} + \mathcal{O}(\epsilon^2)
	\end{equation}
	where $\xi_{b}=\frac{1}{2}\bar{\nabla}_b h^{c}{}_{c}-\bar{\nabla}_c h^{c}{}_{b}$ and $\bar{\Delta}_L$ is the Lichnerowicz operator $\bar{\Delta}_L h_{ab}=-\bar{\nabla}^2 h_{ab}-2\bar{R}_{a}{}^{c}{}_{b}{}^{d}h_{cd}$ computed using $\bar{g}_{ab}=g_{ab}(t^\star)$. When the background is Euclidean space, $\bar{g}_{ab}=\delta_{ab}$,  one recovers a heat equation with an additional contribution in the form of a diffeomorphism which vanishes for pure gauge perturbations, i.e., $h_{ab}=\partial_{(a} u_{b)}$. For a non-trivial Ricci-flat fixed point, the flow may develop curvature singularities in finite time as evidenced by a squashed sphere \cite{topping_2006}. Perelman developed a rich set of methods to surgically remove these singularities which eventually lead to the proof of the Poincar\'e conjecture \cite{2002math.....11159P}. Further results establish the short-time existence of the flow which then allows the geometry to be analytically or numerically evolved (see \cite{topping_2006,Headrick:2006ti,Figueras:2011va}).

	\section{Einstein-Ricci Flows}
	Just as the heat equation can be interpreted as a gradient flow problem for the Dirichlet energy functional $E[f] =\frac{1}{2}\int_\mathcal{M}\abs{\nabla f}^2\sqrt{g}\dd^d x$ with respect to the inner product $\expval{f_i,f_j} = \int_\mathcal{M}f_if_j\sqrt{g}\dd^dx$ in Euclidean space, the Ricci flow, Eqn.~\eqref{eqn:RicciflowEqn}, also admits a similar formulation. In this work, we first review the Ricci flow for the Euclidean Einstein-Hilbert functional. As discussed earlier in \cite{Headrick:2006ti,topping_2006,2002math.....11159P,2003math......3109P}, the flow is ill-behaved for Riemannian metrics. However, the main observation of this paper is to show that under a Wick rotation of the Lorentzian Einstein-Hilbert action, Eqn.~\eqref{eqn:EHaction}, the flow is well-behaved. We will also generalize our discussion to include minimally-coupled\footnote{In fact, the $\mathcal{W}$ and $\mathcal{F}$ entropy functionals considered by Perelman \cite{2002math.....11159P,topping_2006} are examples of non-minimally coupled matter configurations which reproduce the Ricci flow problem up to diffeomorphisms.} matter contributions.

	\subsection{Riemannian Einstein-Ricci Flows}
	To motivate the Ricci flow for Eqn.~\eqref{eqn:EHaction}, consider the variation of the Einstein-Hilbert action $\mathcal{A}[g_{ab}(t)]$ for \textit{Riemannian} manifolds without boundary. Upon integrating a total divergence, one finds 
	\begin{equation}\label{eqn:ERF}
		\dv{t}\mathcal{A}[g(t)] = \frac{1}{2\kappa}\int_\mathcal{M}\qty(R_{ab}-(1/2)R g_{ab})\dot{g}^{ab}\sqrt{{g}}\, \dd^d x.
	\end{equation}
	In order to understand this as a gradient flow, we consider $\mathcal{A}$ to be a Dirichlet energy functional on the space of metrics with an inner product $\expval{\dot{g}_i,\dot{g}_j} = \int_{\mathcal{M}}\dot{g}_{ab}\dot{g}^{ab}$. We wish to find a flow such that the Dirichlet energy converges onto its stationary points. This can be achieved for the Einstein-Ricci flow 
	\begin{equation}
		\dot{g}_{ab}(t) = -2R_{ab}(t) + R(t)g_{ab}(t),\hspace{.2in}g_{ab}(0) = {g}^i_{ab}.
	\end{equation}
	Evaluating the action along this flow, one finds that the Dirichlet functional decreases
	\begin{equation}
		\dot{\mathcal{A}}[g(t)] = -\int_\mathcal{M} \dot{g}_{ab}\dot{g}^{ab} \sqrt{{g}}\dd^{d}x\le 0
	\end{equation}
	where we have set $\kappa=1$. At the stationary points $t^\star$ of the Einstein-Ricci flow, Eqn.~\eqref{eqn:ERF}, one recovers the {Riemannian} Einstein field equations $R_{ab}-(1/2)R g_{ab}|_{t^\star}=0$. Although the stationary points of the Ricci flow and the Einstein-Ricci flow are identical, i.e., Ricci flat spaces, the Einstein-Ricci flow behaves badly from the perspective of PDE theory. If we consider an analogous linearization around the Ricci flat fixed point, we would find linear corrections of the form
	\begin{equation}
		\dot{h}_{ab} = -\bar{\Delta}_\text{L}h_{ab}+\bar{g}_{ab}\qty(\bar{\nabla}_c\bar{\nabla}_dh^{cd}-\bar{\nabla}^2h_c{}^c)+2\bar{\nabla}_{(a}\xi_{b)}.
	\end{equation}
	For Euclidean spaces $\bar{g}_{ab}=\delta_{ab}$, the perturbations can be decomposed to parts where the corrections from the scalar curvature vanish and a part with Weyl scalings $h_{ab}(t)=f(t)\delta_{ab}$ which can be shown to evolve with a diffusion constant $(2-d)$ \cite{topping_2006,Headrick:2006ti}. Hence, the flow may then be expressed as $\dot{f} = (2-d)\nabla^2 f+\cdots$ which is a backwards heat equation. For the Einstein-Ricci flow to be well posed, one either must consider $t\to-t$ or change the inner product in Riemannian signature to counteract the overall signature. Either prescription causes the {Riemannian} Dirichlet functional to increase along the flow.

	\subsection{Lorentzian Einstein-Ricci Flows}
	As we have shown, the Einstein-Hilbert action for Riemannian manifolds\footnote{This statement holds even if the manifold has a boundary.} has an ill-posed Einstein-Ricci flow problem. We show the same is not true for Lorentzian spacetimes which are brought over into the Euclidean section via Wick rotation. One may worry that Lorentzian spacetimes must be static in order to have a well defined Euclidean section. However, the procedure of performing complex deformations on Lorentzian metrics, rather than on Lorentzian time, is now well understood (see \cite{Visser:2017atf} and references therein). One starts with the path integral for the Lorentzian Einstein-Hilbert action
	\begin{equation}
		\mathcal{Z}_L =\int e^{\mathrm{i}\mathcal{A}_L}\mathcal{D}g = \int \exp\qty(\frac{\mathrm{i}}{2}\int_\mathcal{M} R_L\sqrt{-g_L}\dd^{d}x)\mathcal{D}g_L
	\end{equation}
	where $\mathcal{D}g_L$ is the measure on the space of Lorentzian metrics $\mathfrak{G}_L$  and $\mathcal{M}$ is time-orientable with the Lorentzian volume form being $\sqrt{-g_L}\dd^{d}x$. In general, the path integral is not positive definite and oscillates rapidly due to the phase $e^{\mathrm{i}\mathcal{A}_L}$. However, that $\mathcal{M}$ is time-orientable is equivalent to the existence of a non-vanishing timelike vector field $V$ \cite{Visser:2017atf,Hawking:1973uf}. Hence, the Lorentzian path integral can be analytically continued into the Euclidean section via deformations $g_\varepsilon=g+\mathrm{i}\varepsilon (V\otimes V/g(V,V))$ as noted by Visser \cite{Visser:2017atf}. Choosing a contour which avoids the degenerate pole at $\varepsilon=\mathrm{i}$, one recovers the Euclidean path integral (partition function)
	\begin{equation}
		\mathcal{Z} = \int e^{\mathcal{A}_{D}}\mathcal{D}g = \int\exp(-\frac{1}{2}\int_\mathcal{M} R\sqrt{g}\dd^{d}x)\mathcal{D}g.
	\end{equation}
	where we are once again in the space of Riemannian metrics $\mathfrak{G}$ with measure $\mathcal{D}g$.

	It is worth noting that the partition function is not bounded either above or below as conformal fluctuations can be made arbitrarily large \cite{Hawking:1978jn,Hawking:1980gf,Mazur:1989by,Gibbons:1994cg}. However as the semiclassical approximation, near the most dominant saddle point, does reproduce the thermodynamics of black holes, the Euclidean path integral approach does have enough merit to warrant consideration of the deformed Einstein-Hilbert action $\mathcal{A}_{D}$ as a Dirichlet functional. One may object that the Dirichlet functional should be positive definite for a well defined variational problem, so we restrict our considerations to spaces with non-positive scalar curvature $R\le0$. This is quite natural to do from the perspective of black hole thermodynamics in AdS and Minkowski spacetimes.

	For our purposes, we consider $\mathcal{A}_{D}$ as the Dirichlet functional for the Ricci flow problem. Upon first variation, we find
	\begin{equation}\label{eqn:ERF}
		\dot{\mathcal{A}}_D[g(t)] = -\frac{1}{2}\int_\mathcal{M}\qty(R_{ab}-(1/2)R g_{ab})\dot{g}^{ab}\sqrt{{g}}\, \dd^d x.
	\end{equation}
	Then using standard inner product norm $\expval{\dot{g}_i,\dot{g}_j} = \int_{\mathcal{M}}\dot{g}_{ab}\dot{g}^{ab}\sqrt{g}\dd^dx$, the Dirichlet functional decreases $\dot{\mathcal{A}}_{D}\le0$ along the flow
	\begin{equation}\label{eqn:LERF}
		\dot{g}_{ab}(t) = 2 R_{ab}(t)-R(t)g_{ab}(t),\hspace{.2in}g_{ab}(0)={g}^i_{ab}.
	\end{equation}
	Henceforth, we refer to the PDE, Eqn.~\eqref{eqn:LERF}, as the Lorentzian Einstein-Ricci flow. We observe $(i)$ one recovers Einstein gravity at the fixed points of the flow, and $(ii)$ linearizations around the stationary point lead to 
	\begin{equation}
		\dot{h}_{ab} = \bar{\Delta}_\text{L}h_{ab}-\bar{g}_{ab}\qty(\bar{\nabla}_c\bar{\nabla}_dh^{cd}-\bar{\nabla}^2h_c{}^c)-2\bar{\nabla}_{(a}\xi_{b)}
	\end{equation}
	yielding a positive diffusion constant $\dot{f}=(d-2)\nabla^2 f+\cdots$ and a well-posed heat equation for Weyl scalings. To our knowledge, the Lorentzian Einstein-Ricci flow is the first to simultaneously minimize the Euclidean gravity functional and remain well-posed in forward flow time for Ricci-flat spaces.
 
	As an example, consider the homothetic evolution of Einstein spaces $R_{ab}(t)=\frac{2\Lambda}{d-2}\bar{g}_{ab}$. One can check that the exact solution to the Lorentzian Einstein-Ricci flow is
	\begin{equation}
		g_{ab}(t) = \qty[(2/d)+(1-(2/d))e^{-\frac{2d\Lambda}{d-2}t}]\bar{g}_{ab}.
	\end{equation}
	Again, Ricci flat spaces are fixed points which are unchanged along the evolution whereas now negatively curved spaces $(\Lambda<0)$ expand exponentially. Although we consider only non-positive scalar curvature, it is worth noting the finite time singularity of the Ricci flow mentioned earlier is not observed for positively curved spaces $(\Lambda>0)$. 

	\subsection{Generalizations}
	The Lorentzian Einstein-Ricci flow presented thus far is not unique. Indeed, we could add terms which respect the contracted Bianchi identity $\nabla_a(R^{ab}-(1/2)Rg^{ab})=0$ which would only shift the fixed points structure of the flow \cite{Headrick:2006ti}. Concretely, consider the Lorentzian action 
	\begin{equation}
		\mathcal{I}_{L} = \frac{1}{2}\int_\mathcal{M}\qty(R-2\Lambda - \mathcal{L}_\text{M})\sqrt{-g}\dd^dx
	\end{equation}
	where we restrict the matter content $\mathcal{L}_\text{M}\equiv\mathcal{L}_\text{M}(\Phi^I,\nabla\Phi^I)$ to be well defined in the Euclidean section and fixed along the flow $\dot{\Phi}^I=0$. We can construct a deformed Einstein-Hilbert action $\mathcal{I}_D$ via a Wick rotation. The Lorentzian Einstein-Ricci flow problem for the Dirichlet functional $\mathcal{I}_D$ may be formulated as follows; $\dot{\mathcal{I}}_{D}[g(t)]\le0$ on the standard inner product norm $\expval{\dot{g}_i,\dot{g}_j} = \int_{\mathcal{M}}\dot{g}_{ab}\dot{g}^{ab}\sqrt{g}\dd^dx$ for the flow
	\begin{align}\label{eqn:generalization}
		\dot{g}_{ab}(t)&=2R_{ab}(t)-R(t)g_{ab}(t)+2\Lambda g_{ab}(t)-2T_{ab}(g(t)),\nonumber\\
	 	g_{ab}(0)&=\bar{g}_{ab},
	\end{align}
	where the stress energy is ${T}_{ab} = (2/\sqrt{g})\partial(\sqrt{g}\mathcal{L}_\text{M})/\partial g^{ab}$. A few comments are in order. In the absence of matter, the fixed points are now Einstein spaces rather than Ricci-flat spaces. Furthermore, the additional contribution from the cosmological constant respects the diffusive nature of the flow under a linearization as no kinetic terms are added by Weyl scalings, and the Bianchi identity is manifestly obeyed $\nabla_a(R^{ab}-(1/2)R g^{ab}+\Lambda g^{ab})=0$. In the presence of matter, the flow reduces to $R_{ab}-(1/2)Rg_{ab}+\Lambda g_{ab}|_{t^\star}=T_{ab}|_{t^\star}$, and thus imposes $\nabla_a T^{ab}|_{t^\star}=0$. The conservation of stress-energy is equivalent to the statement $\delta \mathcal{I}_{D}[g(t^\star),\Phi_i]/\delta \Phi_i|_\text{on-shell}=0$. To understand the corrections from the stress-energy, consider a linearization of the stress-energy and the flow around $\Lambda=0, \bar{g}_{ab}=\delta_{ab}$. This yields
	\begin{align}
		T_{ab}(g(t)) =& T_{ab}(\bar{g})+\epsilon \pdv{T_{ab}}{g_{cd}}\eval_{t^\star} h_{cd} + \mathcal{O}(\epsilon^2),\\
		\dot{h}_{ab} =& \bar{\Delta}_\text{L}h_{ab}-\bar{g}_{ab}\qty(\bar{\nabla}_c\bar{\nabla}_dh^{cd}-\bar{\nabla}^2h_c{}^c)-2\bar{\nabla}_{(a}\xi_{b)}\nonumber\\
		&+2\Lambda h_{ab}-2\pdv{T_{ab}}{g_{cd}}\eval_{t^\star} h_{cd}.
	\end{align}
	Hence a Weyl fluctuation, $h_{ab}=f(t)\delta_{ab}$, would satisfy $\dot{f}=(d-2)\nabla^2 f - \mu f+\cdots$ where $\mu = 2(\partial T_{ab}/\partial g_{cd})|_{t^\star}\delta_{cd}\delta^{ab}.$ We can understand these fluctuations $\Delta T_{ab}(t)=T_{ab}(g(t))-T_{ab}(\bar{g})$ as sources $Q_{ab}(t) = \Delta T_{ab}(t)$ in the heat equation. In Euclidean space, matter contributions will source exponential factors of the form $e^{-\mu t}$ which competes with diffusion if $\mu<0$ at $t\gg1$. In general, the details depend on the stress-energy of $\mathcal{L}_\text{M}$ in Lorentzian spacetime and we expect the flow to converge only if $\mu$ is non-negative in Euclidean signature.


	\section{Einstein-Ricci Flows as RG Flows}
	The Ricci flow was originally motivated as an RG flow on the worldsheet for 2d non-linear sigma models \cite{PhysRevLett.45.1057,Headrick:2006ti}. In the context of entanglement entropy and AdS/CFT, shape deformations of entanglement surfaces and geometric flows of Randall-Sundrum surfaces have been shown to reproduce the Ricci flow up to corrections \cite{Jackson:2013eqa,Verlinde:1999xm}.

	In light of this, it is suggestive to consider the Lorentzian Einstein-Ricci flow as an RG flow. We can understand the gradient flow as an RG process by identifying $t$ as a renormalization scale with cutoffs at $t_0\le t\le t^\star$. Then we may write the generalized Lorentzian Einstein-Ricci flow somewhat suggestively as
	\begin{align}
		\dot{\Phi}^I &= \beta^I,\label{eqn:RGE1}\\
		\dot{g}_{ab} &= \beta_{ab} + 2\Lambda g_{ab} - 2 T_{ab}.\label{eqn:RGE2}
	\end{align}
	where $\beta_{ab} = 2 R_{ab} - R g_{ab}$ and $\beta^I$ are the beta functions for the matter. As the flow connects configurations $[g(t_0),\Phi^I(t_0)]$ to $[g(t^\star),\Phi^I(t^\star)]$, we construct an off-shell partition function ${Z}_t[g,\Phi]$ along the flow
	\begin{equation}
		Z_t[g,\Phi] = \int \exp(\mathcal{A}^\text{eff}_t[g,\Phi])\mathcal{D}g\mathcal{D}\Phi\eval_{t_0\le t\le t^\star}
	\end{equation}
	where we have integrated out modes with Dirichlet energies outside of the domain of $t$. As the first variation of the action may be written as
	\begin{equation}
		\dv{\mathcal{A}_t^\text{eff}}{t} = \qty(\dot{\Phi}^I\fdv{}{\Phi^I} +\dot{g}_{ab}\fdv{}{g_{ab}})\mathcal{A}_t^\text{eff}.
	\end{equation}
	One finds the relation, upon using Eqn.~\eqref{eqn:RGE1}--\eqref{eqn:RGE2},
	\begin{align}
		\dv{Z_t}{t} &= \qty(\beta^I\fdv{}{\Phi^I} +\qty(\beta_{ab}+2\Lambda g_{ab}-2 T_{ab})\fdv{}{g_{ab}})Z_{t}=0
	\end{align}
	which takes the conventional Callan-Symanzik form. In a usual RG flow, one consider the couplings of the theory to run with the scale. The situation here is qualitatively different as we are considering the deformations of the metric and the matter content along the flow. Essentially, this geometric RG flow deforms an initial metric $g^i$ from the UV stating point to a smoother final configuration at the IR fixed point by virtue of ``coarse graining'' inhomogeneities. This perspective has been espoused thoroughly in the context of AdS/CFT subject to further geometric restrictions \cite{Jackson:2013eqa}. We suspect their analysis can be extended to the generalized Lorentzian Einstein-Ricci flows considered here.

\section{Conclusions}
	We have presented an analysis of the Ricci flow problem for the Lorentzian Einstein-Hilbert action and its generalization in Eqn.~\eqref{eqn:generalization}. This was done by an analytic continuation into the Euclidean section and constructing a gradient flow problem for the Dirichlet functional $\mathcal{I}_D$. For linearizations around Euclidean space and in the absence of matter, the flow is well-posed converging onto Einstein spaces at the fixed points. 

	Several extensions of our analysis are available. One could include non-minimally coupled, dynamical matter configurations to the Lorentzian action with higher order curvature interaction and study the resulting flow equations. Furthermore, it may be useful to study flows of Euclidean black hole spacetimes with conserved charges in the Wald-Iyer formalism. In the context of AdS/CFT, the geometric flows considered here should be dual to RG flows in the boundary CFT \cite{Shyam:2018sro}. As such, it should be possible to study the geometric flow of Euclidean AdS black holes with conserved R-charge as a corresponding RG flow for $\mathcal{N}=4$ SYM at finite temperature on $S^3$ with a $U(1)$ chemical potential. Two questions, which are related, pertain to the Hawking-Page phase transition and the interpretation of the flow time as a spacetime dimension. In \cite{Verlinde:1999xm,Shyam:2018sro}, the radial direction of AdS$_5$ was argued as a flow time and identified with the RG scale. The corresponding RG equations had a similar form to Eqn.~\eqref{eqn:RGE1}--\eqref{eqn:RGE2} in the absence of stress-energy. We know that the confinement/deconfinement phase transition in $\mathcal{N}=4$ is dual to a Hawking-Page phase transition in the bulk \cite{Witten:1998zw}. Hence placing strong RG cutoffs in AdS will screen the black hole spectrum and thus keep the CFT confining. It is natural to wonder how the confinement/deconfinement phase transition may be evaded based on RG scalings and if the flow parameter may be related to thermodynamic parameters. Recently, we have shown that evading the deconfinement/confinement phase transition relied on deforming the R-charge to imaginary values \cite{Cherman:2020zea}. In the context of this paper, we suspect the flow parameter $t$ may be related to the imaginary R-charge.

\begin{acknowledgments}
	\label{sec:acknowledgments}
	\begin{center}
		\textbf{Acknowledgments}
	\end{center}
	I would like to thank Niayesh Afshordi, Jos\'e Tom\'as G\'alvez Ghersi, Soham Mukherjee, Thanu Padmanabhan, Matthew Robbins, Vasudev Shyam, Rafael Sorkin, and Apoorv Tiwari for their valuable comments during the inception of this paper. I appreciate conversations with Jiaping Wang for the present version of this paper. Research at the Perimeter Institute is supported by the Government of Canada through the Department of Innovation, Science and Economic Development Canada. AD is supported by the National Science Foundation Graduate Research Fellowship Program under Grant No.\ 00039202 and by the University of Minnesota.
	\end{acknowledgments}

	\bibliography{ERFEV0.bib}

\end{document}